\documentclass{PoS}
\usepackage{amssymb}
\usepackage{amsmath}
\usepackage{booktabs}

\title{Topological susceptibility and axial symmetry at finite temperature}

\ShortTitle{Topological susceptibility and axial symmetry at finite temperature}

\author{JLQCD Collaboration: \speaker{Guido Cossu}$^a$\thanks{E-mail: cossu@post.kek.jp}, Sinya Aoki$^b$, Shoji Hashimoto$^{a,c}$, Takashi Kaneko$^{a,c}$, Hideo Matsufuru$^a$, Jun-ichi Noaki$^a$, Eigo Shintani$^d$\\
\\
\llap{$^a$}Theory Center, IPNS, High Energy Accelerator Research Organization
  (KEK), Tsukuba 305-0801, Japan\\
\llap{$^b$}Graduate School of Pure and Applied Sciences, University of Tsukuba, Tsukuba 305-8571, Japan\\
\llap{$^c$}School of High Energy Accelerator Science, The Graduate University for
Advanced Studies (Sokendai), Tsukuba 305-0801, Japan\\
\llap{$^d$}RIKEN-BNL Research Center, Upton, NY 11973-5000, USA}

\abstract{We consider the simulation of finite temperature QCD with two flavors of dynamical overlap fermions in order to study the suppression of the axial $U(1)$ symmetry breaking at the chiral phase transition point. As a preliminary study, pure gauge simulations are performed to investigate how fixing the topology affects physical quantities like the topological susceptibility, $\chi_t$, at finite temperature, showing that it is possible to reconstruct known results from the fixed topology sector. First results on the degeneracy of meson correlators in the high temperature QGP sector are shown.}

\FullConference{XXIX International Symposium on Lattice Field Theory \\
		 July 10 - 16 2011\\
		 Squaw Valley, Lake Tahoe, California}

\begin{document}

\section{Introduction}

It is an interesting and long standing problem to understand whether the flavour-singlet axial $U(1)$ symmetry is restored or not above the finite temperature transition in the chirally symmetric phase of QCD. At low temperature it is well known that the chiral symmetry is spontaneously broken while the axial $U(1)$ symmetry  is broken at quantum level by the presence of configurations with non-trivial topological structure. The semi-classical configurations that contribute to the axial charge are called instantons and it can be shown that zero modes of the Dirac operator are related to the presence of these and other configurations that have non-trivial topological charge. Hence, topology plays a major role in this subject. One clear example is the Witten-Veneziano relation \cite{Witten:1979vv,Veneziano:1979ec} that connects the topological susceptibility of pure gauge theory to the heavy mass of the $\eta^\prime(958)$ particle, the candidate to be the would-be Nambu-Goldstone boson of the axial $U(1)$ symmetry. 

Our purpose is to study the behavior of meson correlators at high temperature, looking for signals of (approximate) degeneracy in all the singlet and triplet channels ($\sigma, \delta, \pi, \eta^\prime$) toward the chiral limit. This would be an evidence of effective restoration of both chiral and axial symmetries, several intermediate steps are to be considered to control the systematic errors. The best available discretization of the Dirac operator is the Overlap fermions \cite{Neuberger:1997fp}. Dynamical simulations with Overlap fermions are possible with current machines and were performed by JLQCD collaboration in the past years \cite{Kaneko:2006pa}. The price payed in order to be able to have a very precise fulfillment of the Ginsparg-Wilson relation \cite{Ginsparg:1981bj} on the lattice is to fix the topology. Changing topology in a HMC trajectory requires generating rough configurations associated with near-zero modes of the hermitian Wilson operator ($H_W$). A discontinuity arises in the HMC when one of these modes crosses zero. Treating this discontinuity is a costly task, although it could be achieved using the reflection/refraction methods \cite{Egri:2005cx}. One solution to this problem is to suppress near zero modes of $H_W$ and to avoid the zero crossing, with the disadvantage of preventing topology change during the simulation. The simulation is no more ergodic, and some technique has to be developed to obtain the physical results at $\theta = 0$. This program was indeed developed by JLQCD collaboration (see \cite{Aoki:2007ka} for details) and it works nicely at zero temperature. In order to accomplish our main target, we need to be sure that the same methods can be applied even in the finite temperature case. The first step towards the understanding and controlling the effect of fixing topology at finite temperature is to reproduce the known results of pure gauge theory. In particular we focus on the most relevant quantity, i.e. the topological susceptibility $\chi_t$, and see if it can be reproduced with the same methods adopted at zero temperature. 

The rest of the paper consists of two main sections. The first is devoted to the discussion of our study of topological susceptibility in pure gauge theory at finite temperature. The  second is the measurements of meson correlators, in flavor singlet and triplet channels, and the spectral density of the Overlap Dirac operator, where we show some evidences that the axial $U(1)$ symmetry breaking is suppressed after the chiral phase transition.  In the last section we will draw some conclusions.

\section{Pure gauge simulations}

In a previous paper \cite{Aoki:2007pw}, the JLQCD-TWQCD collaboration calculated the topological susceptibility at zero temperature in the case of two flavors of overlap fermions. They demonstrated that it is possible to obtain physical quantities even with simulations at fixed topological sector. They checked the prediction of $\chi$PT, i.e. $\chi_t = (m_q \Sigma)/N_f + O(m_q^2)$, finding the expected linear behavior of $\chi_t$ versus the sea quark mass. The methods described in  \cite{Aoki:2007ka} were applied in that case, i.e. the topological susceptibility is measured by the long distance value of disconnected correlators of pseudoscalar operators (they are non-zero because in a fixed topology environment clustering property of field theory is violated). We will investigate numerically whether we can apply the same methodology at finite temperature too, with a preliminary test in pure gauge simulations, in order to be trained in dealing with this systematic error in the most interesting full QCD case.

The correlators are calculated assuming that the major contribution comes solely from the lowest 50 eigenmodes. We tested for one value of $\beta$ that this is really the case by changing the number of eigenmodes to 30 and 40. The result is that the long distance behavior saturates at 40-50 eigenvalues, the short distance correlator has a worse approximation, as expected.

In the pure gauge theory,there is an anomalous contribution to the pseudo-scalar meson correlator in the flavor singlet channel, which is called hairpin diagram \cite{DeGrand:2002gm,Bardeen:2000cz}:
\begin{equation}
H(p) = f_P \frac{1}{p^2 + m_\pi^2}m_0^2\frac{1}{p^2 + m_\pi^2}f_P.
\end{equation}
This is a zero temperature result, and we assume that it is valid at finite temperature with all parameters depending on temperature. We actually find that this fits our data very well. Also, the results are independent of the valence quark mass, as it should be in the case of pure gauge theory. For statistical purposes, results are obtained with a joint fit of connected and disconnected part. The joint fit assumes an identical decay rate for the two correlators, long distance of connected part going to zero, to constrain the rest of parameters. Fits also discard the first and last three points of the spatial separation range. 

Simulations were performed using an Iwasaki lattice gauge action, on a lattice of dimensions $24^3\times 6$ and $\beta \in [2.35, \dots, 2.55]$ in order to be in the range of temperatures $[249, \dots, 347]$ MeV. The phase transition was estimated by the Polyakov loop behavior to be at $T=288$ MeV. Except for one run at $Q=1$ all configurations were generated at $Q=0$.

\begin{figure}[t]
  \centering
  \includegraphics[clip=true, width=0.5\columnwidth]{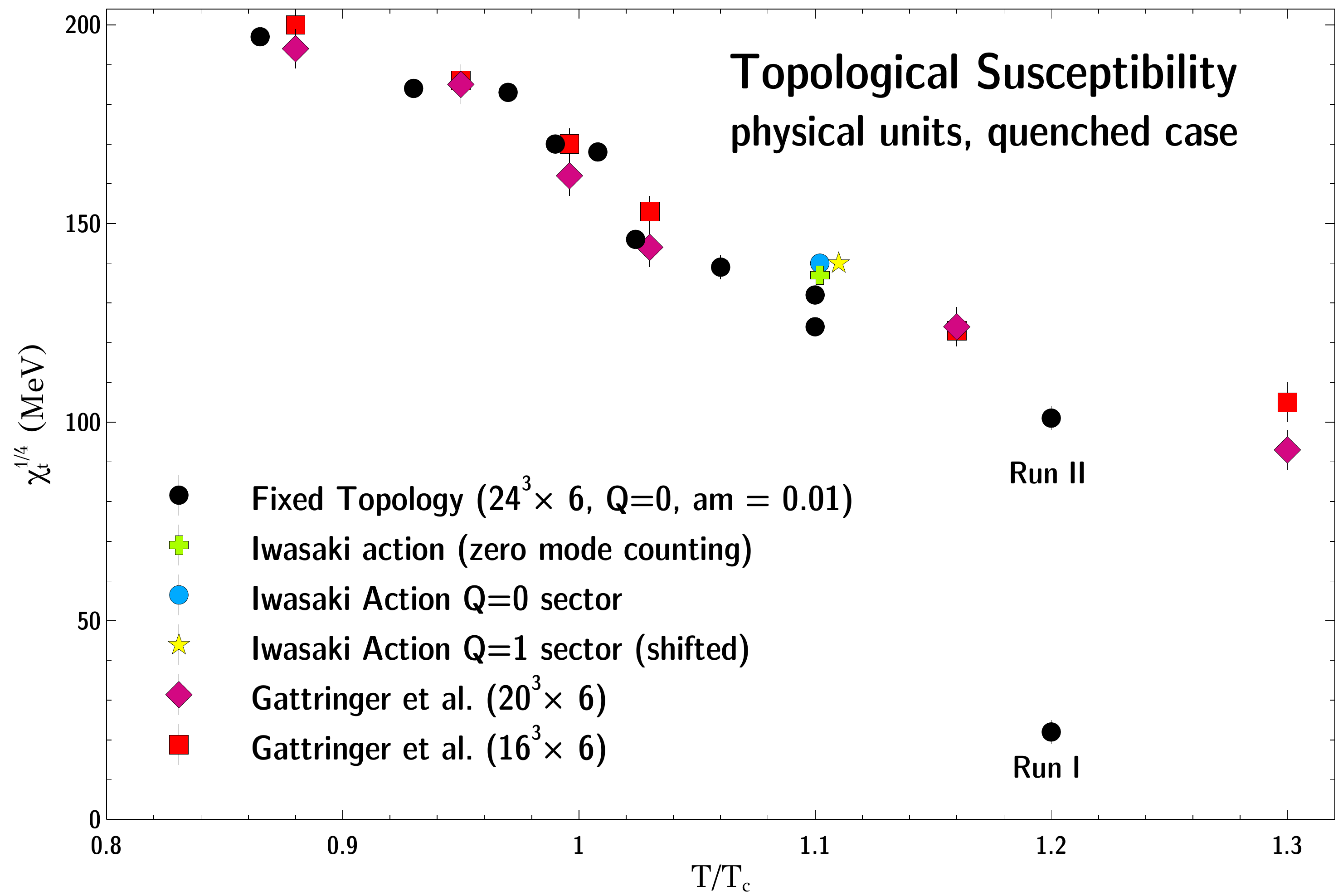}
  \caption{(Topological susceptibility on the $24^3\times 6$ lattice.  Statistical errors only. Reference data points are diamonds and squares. The black dots are our results from measurements at Q=0.}
  \label{fig:TopSusc} 
\end{figure}

The outcome for the topological susceptibility measurements is shown in figure~\ref{fig:TopSusc}. Our results are shown as black dots. They follow with a good accuracy the reference results by the Regensburg group \cite{Gattringer:2002mr}, obtained with direct zero eigenmode counting. In order to cross-check the methodology we also accumulated configurations without fixing topology, by using the same HMC algorithm. The parameters for this run were chosen so that it corresponds to the same temperature as the $\beta = 2.50$ run, $T/T_C \simeq 1.1$. We then extracted two subsets of configurations with $Q=0$ and $Q=1$ and applied the correlator methodology to extract topological susceptibility. The final results agree among two sectors and moreover are in accordance with an estimate using the Atiyah-Singer's index theorem for all configurations (points indicated by cross $Q=$all, blue dot $Q=0$ and star $Q=1$ in the figure). This result indicates that the term in the action that forces fixed topology is not preventing appearance of local topological fluctuations, like pairs of caloron-anticalorons (so called topological molecules) which are associated to very low eigenmodes. These topological fluctuations are distributed in the same way as the ones generated without fixing topology. 

\begin{figure}[ht]
 \centering
   \includegraphics[clip=true, width=0.7\columnwidth]{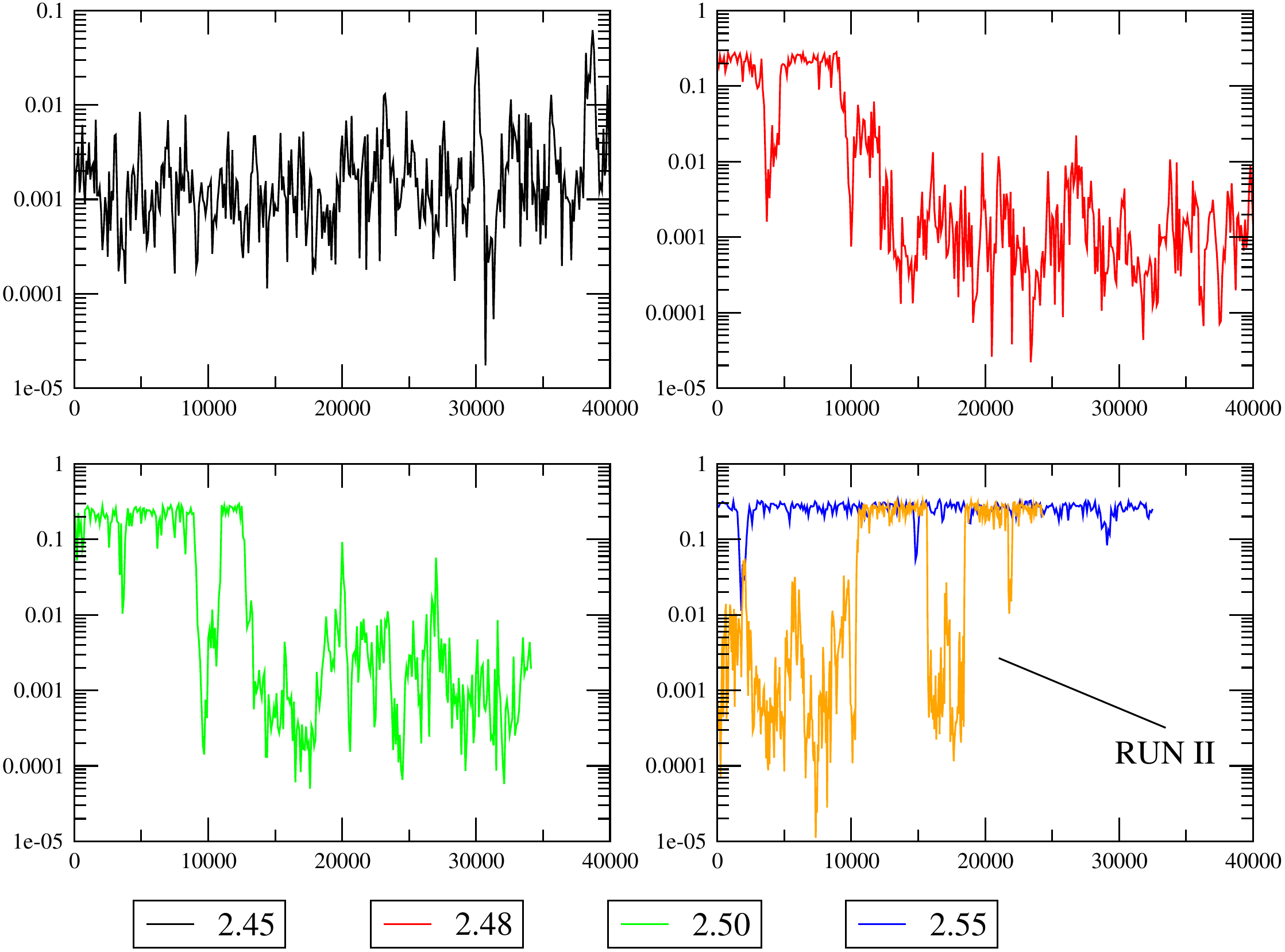}
  \caption{Lowest eigenvalues at several temperatures. On abscissas the HMC trajectory number, one point every 100 trajectories except for RUN-II where we saved one configuration every 50.  }
  \label{fig:Lowest} 
\end{figure}

So far we demonstrated that our method works very well even in the finite temperature case. However we discovered and reported at the Lattice conference that topology fixing poses potential problems in the high temperature regime, at finer lattice spacing. The lowest eigenmodes (main source of the topological susceptibility signal) take longer time to thermalize than usual thermodynamical quantities. This is a known problem but in our case the effect appeared more strongly. The points at $\beta = 2.50$ and $\beta = 2.55$ were the most problematic, exhibiting some dependence on the initial configuration, see Figure~\ref{fig:Lowest}. In particular the run tagged RUN-II, obtained by starting from a thermalized configuration at $\beta = 2.50$ and heated up to $\beta = 2.55$, was giving different results from a similar run starting from unit configuration. This was a severe problem, mining the reliability of higher temperature runs. We thus further investigated the issue by accumulating more configurations, especially for the highest temperature point. Final result is shown in Figure~\ref{fig:Lowest}. We recall that all runs started from a unit configuration, high temperature, and then thermalized (except RUN-II). Thermalization is monitored by plaquette and Polyakov Loop as standard observables. By looking at the lowest eigenvalue history, the first three panels do not show any particular problem in thermalization, except the expected increasing autocorrelation time going to finer lattices. With newest data no anomalous behavior is detected also for $\beta = 2.55$. At Lattice 2011 the history was much shorter, less than 10k trajectories, and no jumps were present. Transition between two different values of the lowest eigenmode is sharp but the tunneling rate is non-zero, and no dependence on the starting configuration is observed. The issue was just statistical and is not expected to affect full QCD results.

We thus conclude that physical quantities can be obtained from fixed topology simulations at finite temperature.

\section{Dynamical overlap simulations}

In this section we will discuss some of the results from simulations of two flavors of dynamical overlap fermions in the high temperature region.

We accumulated O(200-300) thermalized configurations per $\beta$ with several masses and temperatures, see table~\ref{table:DynFerm}. Accumulating the corresponding zero temperature configurations would have been a really demanding task so we do not have currently a measurement of the lattice spacing and pion mass for every $\beta$. At $\beta = 2.30$ and bare quark mass $am=0.015$ was 286 MeV, 360 MeV at $am = 0.025$ \cite{Aoki:2008tq}. By looking at the Dirac operator spectral density we observe that a gap is already opened at a temperature of around 192 MeV ($\beta = 2.25$) near the chiral limit, $am = 0.01$, confirmed by higher temperature simulations. This could be a signal of strong suppression of axial symmetry breaking, because the splitting between flavor-singlet and non-singlet correlators dominantly comes from the near-zero modes, as we demonstrate below. We do not observe this behavior with current data at the temperature of 177 MeV ($\beta = 2.20$,  $am > 0.025$).

\begin{figure}[t]
  \centering
  \includegraphics[clip=true, width=0.41\columnwidth]{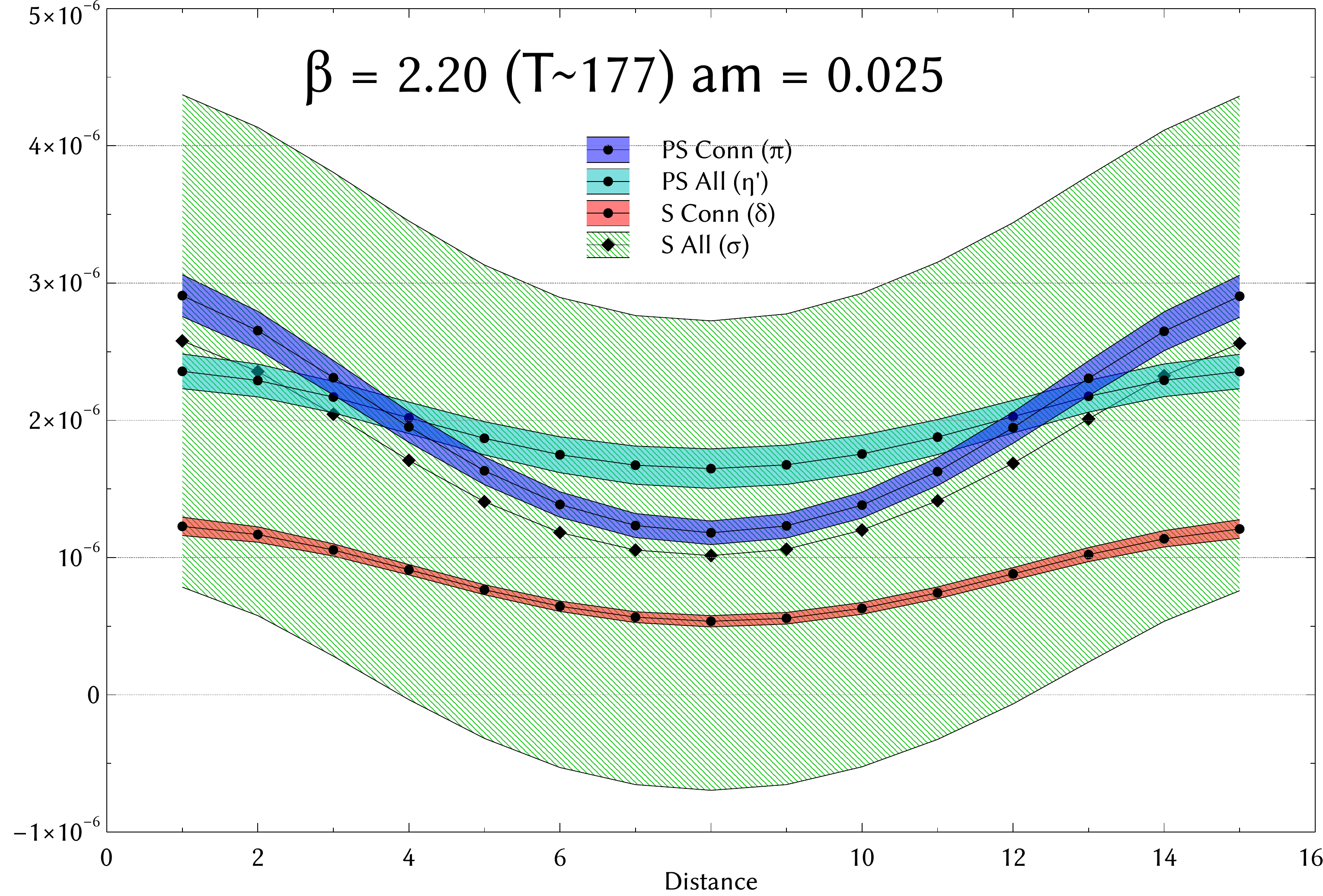}
  \includegraphics[clip=true, width=0.41\columnwidth]{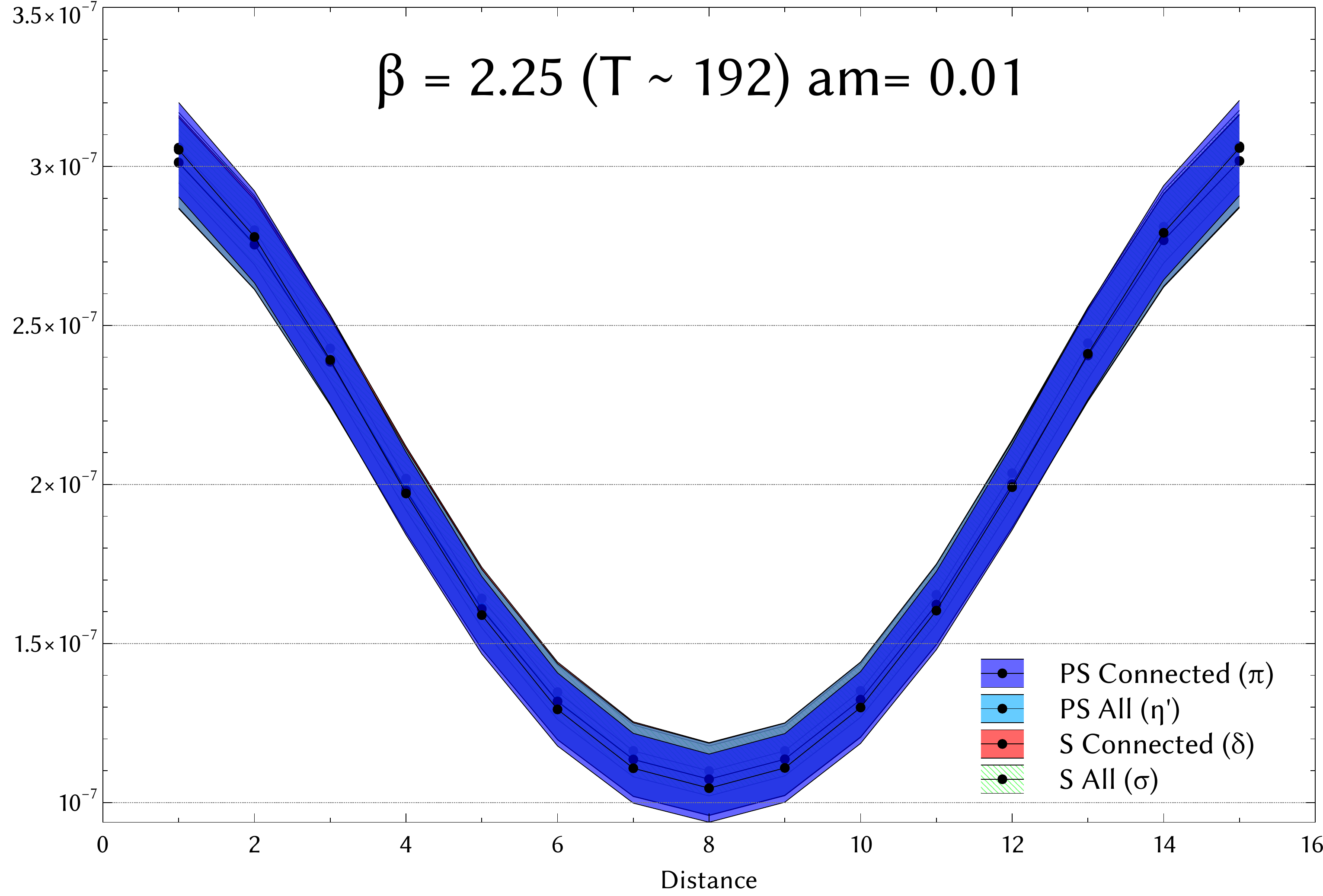}
  \\
  \includegraphics[clip=true, width=0.41\columnwidth]{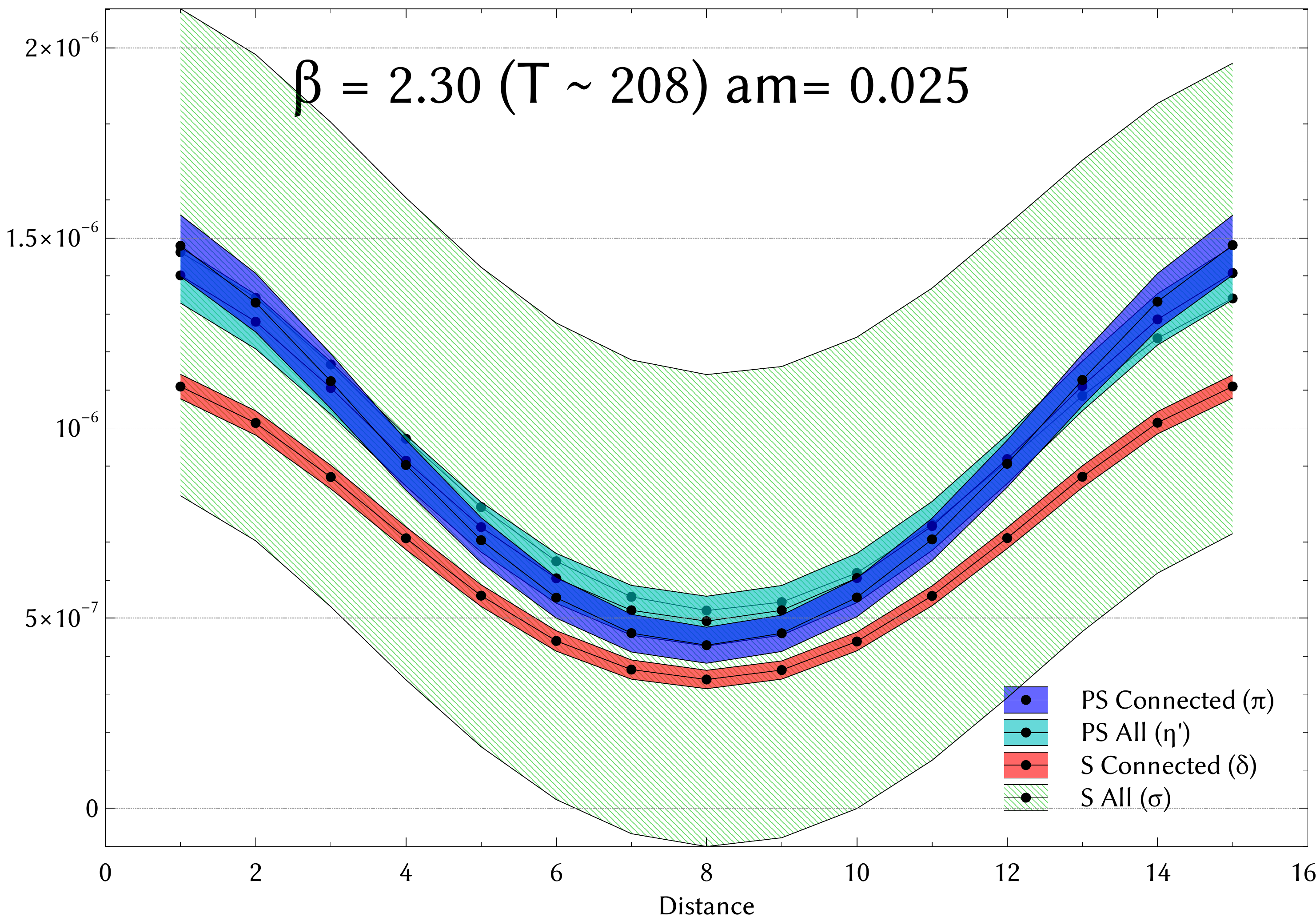}
  \includegraphics[clip=true, width=0.41\columnwidth]{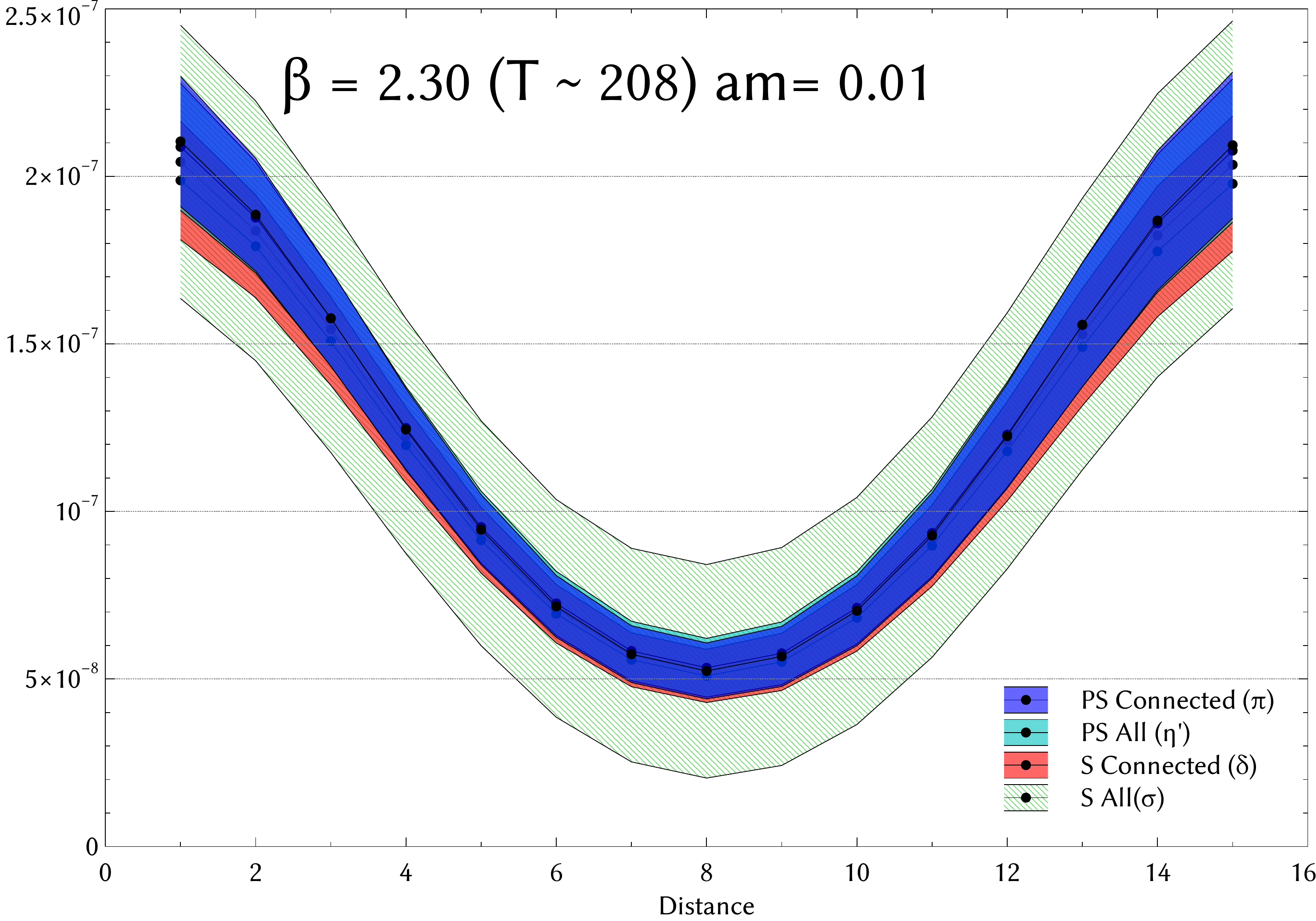}
  \caption{Meson correlators at several temperatures and masses.}
  \label{fig:Mesons_Nf2} 
\end{figure}

 We report also our first observation of the meson correlators in this range of temperatures (figure~\ref{fig:Mesons_Nf2}). We plot the singlet and non-singlet pseudo-scalars ($\eta^\prime$ and $\pi$) and singlet and non-singlet scalars ($\sigma$ and $\delta$).

Data at $\beta = 2.20$ do not show any degeneracy among the correlators of the lightest mesons. This could be an effect of the high mass ($am=0.025$) but it is also possible that this temperature is still in the chirally broken region (see the spectral density plot). The number of configurations is not sufficient to precisely determine the transition temperature, so we cannot settle this question. The meson correlators at the highest temperature, starting from $\beta = 2.25$, show a clear tendency toward degeneracy in the chiral limit, a signal that $U(1)_A$ axial symmetry is effectively restored.

We need to accumulate more data in order to be conclusive, but certainly we observe robust signals of axial symmetry restoration. Further study around the phase transition is on the way. 

\begin{table*}[b]
  \centering
  \begin{tabular}{ c  c  c c }
    \toprule[0.5pt]
    $\beta$ & $a$(fm) & $T$(MeV) & Masses ($am$)\\
    \midrule[0.2pt]
     2.18  & 0.144   & 171  & 0.05\\
     2.20  & 0.139   & 177  & 0.05, 0.025\\
     2.25  & 0.128   & 192  & 0.01\\
     2.30  & 0.118   & 208  & 0.05, 0.025, 0.01\\
     \bottomrule[0.5pt]
  \end{tabular}
  \caption{\label{table:DynFerm}Parameters of some of the collected data for two flavors of dynamical overlap fermions, lattice dimensions $16^3\times 8$. The reported lattice spacings are the extrapolations to the chiral limit.}
\end{table*}

\begin{figure}[t]
  \centering
  \includegraphics[clip=true, width=0.50\columnwidth]{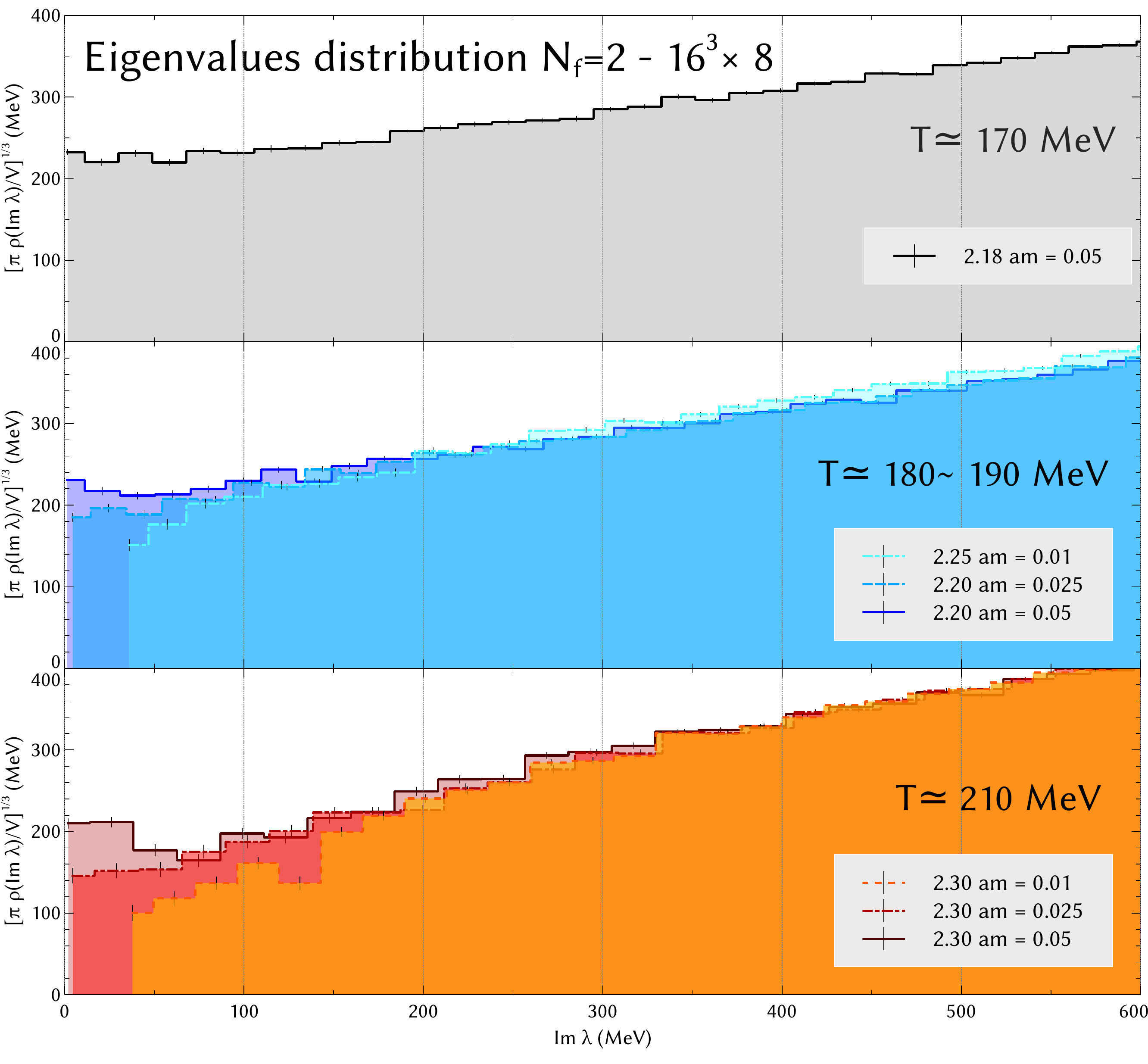}
  \caption{Spectral density of the overlap Dirac operator for $N_f=2$. The several $\beta$s were isolated to emphasize the mass dependence of the density. Zero counting of eigenvalues is intended on the left when line stops.}
  \label{fig:Spect_Dens_Nf2} 
\end{figure}

\section{Conclusions and perspectives}

We have shown that it is possible to perform finite temperature lattice simulations with overlap fermions at fixed topology. It is tested in the case of pure gauge theory where measurements of topological susceptibility can be compared with known results. 

Having under control the systematics of topology fixing we performed simulations of two flavors of overlap fermions at several temperatures in the range $[170\sim 208]$ MeV. We measured the correlators of meson operators in all the (pseudo)scalar channels looking for their degeneracy in the chiral limit, signal of effective restoration of both chiral and axial symmetry. We found evidence of this restoration, corroborated also by the spectral density analysis that exhibits a gap in the chiral limit at temperatures above $>192$ MeV, in the current data set.

We can currently fairly say that we have clear evidence of $U(1)_A$ effective restoration in a region just above the chiral phase transition in two flavors QCD. The next step is narrowing the region of uncertainty about the temperature when the gap starts opening. In a forthcoming paper in preparation a complete analysis and the newly collected data will be presented.
 
This work is supported in part by the HPCI Strategic Program of Ministry of Education and in part by the Grant-in-Aid for Scientific Research on Innovative Areas (No. 2004: 20105001, 20105003, 20105005, 21674002, 21684013).

\bibliographystyle{unsrt}
\bibliography{proceeding.bib}

\begin{thebibliography}{10}

\bibitem{Witten:1979vv}
Edward Witten.
\newblock {Current Algebra Theorems for the U(1) Goldstone Boson}.
\newblock {\em Nucl.Phys.}, B156:269, 1979.

\bibitem{Veneziano:1979ec}
G.~Veneziano.
\newblock {U(1) Without Instantons}.
\newblock {\em Nucl.Phys.}, B159:213--224, 1979.

\bibitem{Neuberger:1997fp}
Herbert Neuberger.
\newblock {Exactly massless quarks on the lattice}.
\newblock {\em Phys.Lett.}, B417:141--144, 1998.

\bibitem{Kaneko:2006pa}
T.~Kaneko et~al.
\newblock {JLQCD's dynamical overlap project}.
\newblock {\em PoS}, LAT2006:054, 2006.

\bibitem{Ginsparg:1981bj}
Paul~H. Ginsparg and Kenneth~G. Wilson.
\newblock {A Remnant of Chiral Symmetry on the Lattice}.
\newblock {\em Phys.Rev.}, D25:2649, 1982.

\bibitem{Egri:2005cx}
G.I. Egri, Z.~Fodor, S.D. Katz, and K.K. Szabo.
\newblock {Topology with dynamical overlap fermions}.
\newblock {\em JHEP}, 0601:049, 2006.

\bibitem{Aoki:2007ka}
Sinya Aoki, Hidenori Fukaya, Shoji Hashimoto, and Tetsuya Onogi.
\newblock {Finite volume QCD at fixed topological charge}.
\newblock {\em Phys.Rev.}, D76:054508, 2007.

\bibitem{Aoki:2007pw}
S.~Aoki et~al.
\newblock {Topological susceptibility in two-flavor lattice QCD with exact
  chiral symmetry}.
\newblock {\em Phys.Lett.}, B665:294--297, 2008.

\bibitem{DeGrand:2002gm}
Thomas~A. DeGrand and Urs~M. Heller.
\newblock {Witten-Veneziano relation, quenched QCD, and overlap fermions}.
\newblock {\em Phys.Rev.}, D65:114501, 2002.

\bibitem{Bardeen:2000cz}
William~A. Bardeen, A.~Duncan, E.~Eichten, and H.~Thacker.
\newblock {Anomalous chiral behavior in quenched lattice QCD}.
\newblock {\em Phys.Rev.}, D62:114505, 2000.

\bibitem{Gattringer:2002mr}
Christof Gattringer, Roland Hoffmann, and Stefan Schaefer.
\newblock {The Topological susceptibility of SU(3) gauge theory near T(c)}.
\newblock {\em Phys.Lett.}, B535:358--362, 2002.

\bibitem{Aoki:2008tq}
S.~Aoki et~al.
\newblock {Two-flavor QCD simulation with exact chiral symmetry}.
\newblock {\em Phys.Rev.}, D78:014508, 2008.

\end{thebibliography}

\end{document}